\documentclass{article}
\usepackage{spconf,amsmath,graphicx}
\usepackage{hyperref}
\usepackage{url,times,booktabs, tabularx}
\usepackage{multirow}
\usepackage{caption}
\usepackage{subcaption}
\usepackage{enumitem}
\usepackage{color}
\usepackage{cleveref}
\usepackage{comment}
\usepackage{amssymb}

\usepackage[numbers, sort&compress, square]{natbib}

\usepackage{siunitx}
\usepackage{xspace}

\newcommand{\etalcite}[1]{ et al\@.~\cite{#1}}
\newcommand*{\gccphat}{\mbox{GCC-PHAT}\xspace}
\newcommand{\aname}[1]{\textit{#1}}

\newcommand{\T}{\mathsf{T}}
\renewcommand{\H}{\mathsf{H}}

\newcommand{\ER}{$\text{ER}_{\le \SI{20}{\degree}}$\xspace}
\newcommand{\Fone}{$\text{F}_{\le \SI{20}{\degree}}$\xspace}
\newcommand{\LE}{$\text{LE}_\text{CD}$\xspace}
\newcommand{\LR}{$\text{LR}_\text{CD}$\xspace}

\renewcommand{\T}{\mathsf{T}}
\renewcommand{\H}{\mathsf{H}}

\makeatletter
\newcommand{\thickhline}{%
    \noalign {\ifnum 0=`}\fi \hrule height 1pt
    \futurelet \reserved@a \@xhline
}
\newcolumntype{"}{@{\hskip\tabcolsep\vrule width 1pt\hskip\tabcolsep}}
\makeatother

\title{SALSA-Lite: A Fast and Effective Feature for Polyphonic Sound Event Localization and Detection with Microphone Arrays}
\name{
    Thi Ngoc Tho Nguyen$^{\star}$, 
    Douglas L. Jones$^{\dagger}$, 
    Karn N. Watcharasupat$^{\star}$,
    Huy Phan$^{\ddagger}$,
    Woon-Seng Gan$^{\star}$
    \thanks{This research was supported by the Singapore Ministry of Education Academic Research Fund Tier-2, under research grant MOE2017-T2-2-060, and the Google Cloud Research Credits program with the award GCP205311440.
    K. N. Watcharasupat acknowledges the support from the CN Yang Scholars Programme, NTU.}
}

\address{
$^{\star}$School of Electrical and Electronic Engineering, Nanyang Technological University (NTU), Singapore\\
$^{\dagger}$Dept. of Electrical and Computer Engineering, University of Illinois at Urbana-Champaign, USA\\
$^{\ddagger}$School of Electronic Engineering and Computer Science, Queen Mary University of London, UK
}

\usepackage{jabbrv}
\DefineJournalException{International Workshop on Machine Learning for Signal Processing (MLSP)}{Proc. MLSP}

\DefineJournalException{IEEE Transactions on Intelligent Transportation Systems}{IEEE Trans. ITS}

\DefineJournalException{Interspeech}{Proc. Interspeech}

\DefineJournalException{IEEE International Conference on Robotics and Automation, 2004. Proceedings. ICRA '04. 2004}{Proc. ICRA}

\DefineJournalException{Proceedings of the Detection and Classification of Acoustic Scenes and Events 2019 Workshop (DCASE2019)}{Proc. DCASE Workshop}

\DefineJournalException{Proceedings of the Detection and Classification of Acoustic Scenes and Events 2020 Workshop (DCASE2020)}{Proc. DCASE Workshop}

\DefineJournalException{2018 IEEE International Conference on Acoustics, Speech and Signal Processing (ICASSP)}{Proc. ICASSP}

\DefineJournalException{International Conference on Acoustics, Speech and Signal Processing (ICASSP)}{Proc. ICASSP}

\DefineJournalException{Proceedings of the IEEE/CVF Conference on Computer Vision and Pattern Recognition (CVPR)}{Proc. CVPR}

\DefineJournalException{2019 IEEE Workshop on Applications of Signal Processing to Audio and Acoustics (WASPAA)}{Proc. WASPAA}

\DefineJournalException{IEEE International Conference on Acoustics, Speech and Signal Processing (ICASSP)}{Proc. ICASSP}

\DefineJournalException{IEEE Transactions on Audio, Speech, and Language Processing}{IEEE Trans. ASLP}

\DefineJournalException{IEEE/ACM Transactions on Audio, Speech, and Language Processing}{IEEE/ACM Trans. ASLP}

\DefineJournalException{Proceedings of the 34th AAAI Conference on Artificial Intelligence}{Proc. AAAI}

\DefineJournalException{Proceedings of the Annual Conference of the International Speech Communication Association}{Proc. Interspeech}

\DefineJournalException{2013 IEEE International Conference on Acoustics, Speech and Signal Processing}{Proc. ICASSP}

\DefineJournalException{Conference Track Proceedings of the 3rd International Conference on Learning Representations}{Proc. ICLR}

\DefineJournalException{International Conference on Machine Learning}{Proc. ICML}

\DefineJournalException{ICASSP}{Proc. ICASSP}

\DefineJournalException{IEEE Transactions on Signal Processing}{IEEE Trans. Signal Process.}

\usepackage{fancyhdr}

\fancypagestyle{firstpage}{
    \fancyhf{}
    \fancyfoot[L]{\footnotesize\textcopyright 2021 IEEE.  Personal use of this material is permitted. Permission from IEEE must be obtained for all other uses, in any current or future media, including reprinting/republishing this material for advertising or promotional purposes, creating new collective works, for resale or redistribution to servers or lists, or reuse of any copyrighted component of this work in other works.}
 }
 
\begin{document}
\ninept

\maketitle

\begin{abstract}
Polyphonic sound event localization and detection (SELD) has many practical applications in acoustic sensing and monitoring. However, the development of real-time SELD has been limited by the demanding computational requirement of most recent SELD systems. In this work, we introduce SALSA-Lite, a fast and effective feature for polyphonic SELD using microphone array inputs. SALSA-Lite is a lightweight variation of a previously proposed SALSA feature for polyphonic SELD. SALSA, which stands for Spatial Cue-Augmented Log-Spectrogram, consists of multichannel log-spectrograms stacked channelwise with the normalized principal eigenvectors of the spectrotemporally corresponding spatial covariance matrices. In contrast to SALSA, which uses eigenvector-based spatial features, SALSA-Lite uses normalized inter-channel phase differences as spatial features, allowing a 30-fold speedup compared to the original SALSA feature. Experimental results on the TAU-NIGENS Spatial Sound Events 2021 dataset showed that the SALSA-Lite feature achieved competitive performance compared to the full SALSA feature, and significantly outperformed the traditional feature set of multichannel log-mel spectrograms with generalized cross-correlation spectra. Specifically, using SALSA-Lite features increased localization-dependent F1 score and class-dependent localization recall by \SI{15}{\percent} and \SI{5}{\percent}, respectively, compared to using multichannel log-mel spectrograms with generalized cross-correlation spectra. 
\end{abstract}


\section{Introduction}
\label{sec:intro}
Sound event localization and detection (SELD) is an emerging research topic that unifies the tasks of sound event detection (SED) and direction-of-arrival estimation (DOAE). SELD as a task involves estimation of the directions of arrival (DOA), the onsets, and the offsets of the detected sound events, while simultaneously classifying their sound classes ~\cite{adavanne2019seld}. Because of the need for source localization, SELD typically requires multichannel audio inputs from a microphone array, whose channel encoding exists in several formats, such as first-order ambisonics (FOA) and generic microphone array (MIC). In this paper, we focus on MIC format, which is the most accessible and commonly-used type of microphone arrays in practice. 

Polyphonic SELD is a relatively new research topic and the majority of recent developments for SELD has been targeted at the model architectures. In 2018, Adavanne\etalcite{adavanne2019seld} pioneered a seminal work that used a convolutional recurrent neural network (CRNN), \aname{SELDnet}, for polyphonic SELD. Cao\etalcite{cao2019twostage} proposed a two-stage strategy with separate SED and DOAE models. Nguyen\etalcite{tho2020smn, tho2021gennet} introduced a \aname{Sequence Matching Network} (SMN) that matched the SED and DOAE output sequences using a bidirectional gated recurrent unit (BiGRU). Cao\etalcite{cao2021ienv2} later proposed a two-branch network, \aname{Event Independent Network} (EIN), that used soft parameter sharing between the SED and DOAE encoder branches, and used multi-head self-attention (MHSA) to output track-wise predictions. Shimada\etalcite{shimada2021accdoa} proposed a unified output representation called \aname{Activity-Coupled Cartesian Direction of Arrival} (ACCDOA) to combine SED and DOAE predictions and losses into a single optimization objective, as well as incorporated a new building block, D3Net, into a CRNN for SELD~\cite{takahashi2021d3net}. 

On the input feature aspect, multichannel magnitude and phase spectrograms were initially used to train SELDnet for polyphonic SELD~\cite{adavanne2019seld}. Subsequently, many works~\cite{cao2019twostage, cao2021ienv2, shimada2021accdoa, politis2020dcasedataset, politis2021dcasedataset, wang2021fouraug} converged on multichannel log-mel spectrograms with spatial channels -- intensity vector (IV) for the FOA format, and generalized cross-correlation with phase transform (\gccphat) for the MIC format -- as de facto standard input features for SELD. We refer to these two features as \textsc{MelSpecIV} and \textsc{MelSpecGCC}, respectively. Raw waveform has also been used for SELD~\cite{he2021sounddet} but is not as popular as other time-frequency features.
Several studies have shown that, using the same network architectures, models trained on the MIC-format \textsc{MelSpecGCC} feature performed worse than those trained on the FOA-format \textsc{MelSpecIV} feature~\cite{politis2020dcasedataset, politis2021dcasedataset, tho2021salsa}. 
When IVs are stacked with spectrograms, the spatial cues in IVs align with the signal energy in the spectrograms along the frequency dimension. This frequency alignment is crucial for networks to resolve multiple overlapping sound events because signals from different sound sources often have their own unique pattern in the frequency axis. On the other hand, the time-lag dimension of the \gccphat features does not have a local one-to-one mapping with the frequency dimension of the spectrograms. As a result, all of the DOA information is aggregated at the frame level, making it difficult to assign correct DOAs to different sound events. 
In addition, \gccphat features are more computationally expensive to compute than IVs. Since the MIC format is the most commonly used type of microphone array in practice, a better SELD feature for the MIC format is needed.

Spatial Cue-Augmented Log-Spectrogram (SALSA) is a recently proposed feature with exact time-frequency (TF) mappings between the signal power and the source directional cues supporting both the FOA and the MIC formats \cite{tho2021salsa}.
However, SALSA is computationally expensive due to the need to compute the principal eigenvectors. In this paper, we propose a computationally cheaper variant of SALSA for the MIC format called SALSA-Lite. In SALSA, the spatial feature is computed from the principal eigenvector of the spatial covariance matrices (SCMs), while in SALSA-Lite, the spatial feature is the frequency-normalized inter-channel phase differences (IPD), which is computed directly from the multichannel complex spectrograms. As a result, SALSA-Lite is significantly more computationally efficient. IPD features are popular for blind source separation~\cite{araki2007bss, traa2013bssipd}, multichannel speech enhancement and separation~\cite{ wang2018multichannelclustering}, and two-channel DOAE~\cite{zhang2010twomicdoa, pak2019ipdenhancementdnn}, which involve implicit or explicit localization using spatial cues. In practice, the spatial cues in IPDs are noisy due to acoustic reverberation, mismatch in the phase responses of the microphones, and the wrap-around effect caused by spatial aliasing \cite{traa2013bssipd}. Therefore, IPD features often require additional enhancements using either a mathematical model~\cite{traa2013bssipd} or a neural model~\cite{pak2019ipdenhancementdnn}. 
For applications in source localization, IPD features are more popular for stereo arrays than higher-order arrays, as there exists more effective localization algorithms such as MUSIC~\cite{schmidt1986multiple} and beamforming~\cite{capon1969mvdr} for the latter. In this study, we show that SALSA-Lite, which is a combination of multichannel linear-frequency log-power spectrograms and a simple normalized version of IPDs, is in fact an effective feature for SELD using microphone array with more than two channels.  

Experimental results on the TAU-NIGENS Spatial Sound Events (TNSSE) 2021 dataset~\cite{politis2021dcasedataset} showed that SALSA-Lite achieved similar performance as SALSA, while boasting a 30-fold speedup. In addition, SALSA-Lite feature significantly outperformed the \textsc{MelSpecGCC} features, which is the de facto standard feature for SELD with microphone arrays, while being \num{9} times faster to compute. This demonstrates SALSA-Lite as a promising candidate feature for real-time SELD using microphone array input signals. 

\section{SELD features for microphone array inputs}
\label{sec:input}


\subsection{Log-spectrograms and \gccphat for MIC format}

For the MIC format, mel-scale \textsc{SpecGCC} feature, which consists of multichannel log-mel spectrograms and pair-wise \gccphat, is arguably the most popular feature for polyphonic SELD. Given an $M$-channel array inputs, the \gccphat is computed for each audio frame $t$ for each of the microphone pairs $(i, j)$ by~\cite{cao2019twostage}
\begin{equation}
    \text{GCC-PHAT}_{i,j}(t, \tau) = \mathcal{F}^{-1}_{f\rightarrow \tau}\left[
        \frac{X_i(t,f) X_j^{\ast}(t,f)}{\|X_i(t,f) X^{\ast}_j(t,f)\|}\right],
    \label{eq:gcc-phat}
\end{equation}
where $\mathcal{F}^{-1}$ is the inverse Fourier transform; 
$t$ and $f$ are the time and frequency indices, respectively; $|\tau|\le f_\text{s} d_\text{max}/c$ is the time lag, where $f_\text{s}$ is the sampling rate, $c\approx\SI{343}{\meter\per\second}$ is the speed of sound, and $d_\text{max}$ is the largest distance between any two microphones; and $X_i(t,f)\in\mathbb{C}$ is the short-
time Fourier transform (STFT) of the $i^{th}$ microphone signal. When the \gccphat features are stacked with mel-scale spectrograms, the ranges of $\tau$ is truncated to $(-K/2, K/2]$, where $K$ is the number of mel bands, resulting in the \textsc{MelSpecGCC} features in $\mathbb{R}^{(M^2+M)/2 \times T \times K}$, 
where
$T$ is the number of time frames.  

\subsection{SALSA}

The $(2M-1)$-channel SALSA feature for the MIC format is formed by stacking the $M$-channel linear-frequency log-power spectrograms with the $(M-1)$-channel eigenvector-based phase vector (EPV). The EPV channels of the SALSA feature approximates the relative distances of arrival (RDOAs) and are computed using
\begin{equation}
    \mathbf{V}(t,f) =-c\left(2 \pi f\right)^{-1}\arg
    \left[U_1^{\ast}(t,f)\mathbf{U}_{2:M}(t,f)\right]\in\mathbb{R}^{M-1},
    \label{eq:iv}
\end{equation}
where $\mathbf{U}(t, f)\in\mathbb{C}^{M}$ is the principal eigenvector of the SCM $\mathbf{R}(t,f) = \mathbb{E}[\mathbf{X}(t,f)\mathbf{X}^{\mathsf{H}}(t,f)]$, where $(\cdot)^{\mathbf{\H}}$ denotes the Hermitian transpose.
To avoid spatial aliasing, elements of $\mathbf{V}$ are zeroed for all TF bins above aliasing frequency. In addition, magnitude and coherence tests are used to select single-source TF bins~\cite{tho2021salsa}. The values of $\mathbf{V}$ are set to zeros for non single-source TF bins.   

\subsection{The proposed feature: SALSA-Lite}
\label{subsec:salsa-lite}

For faster computation, SALSA-Lite replaces the EPV of the SALSA feature with a simple frequency-normalized IPD (NIPD)~\cite{araki2007bss}, which is somewhat equivalent to IV for the FOA format. The NIPD is computed via
\begin{equation}
    \mathbf{\Lambda}(t,f) =-c\left(2 \pi f\right)^{-1}\arg
    \left[X_1^{\ast}(t,f)\mathbf{X}_{2:M}(t,f)\right]\in\mathbb{R}^{M-1}.
    \label{eq:pv}
\end{equation}

To understand the rationale for SALSA-Lite feature, consider the case of a TF bin dominated by a single sound source $S(t, f)$,
\begin{align}
    \mathbf{X}(t, f) \approx \mathbf{H}(t, f, \phi, \theta) S(t, f) + \mathbf{N}(t, f),
\end{align}
where $\mathbf{H}(t, f,\phi, \theta)$ is the theoretical steering vector, $\mathbf{N}(t, f)$ is noise, $\phi$ and $\theta$ are the azimuth and elevation angles, respectively. 
For an $M$-channel farfield array of an arbitrary geometry the theoretical array response can be modelled by 
\begin{equation}
    H_m(t, f, \phi, \theta)
    = \exp\left(-\jmath 2\pi f d_{1m}(t) / c\right),
\label{eq:mic_steervec}
\end{equation} \nobreak
where $\jmath$ is the imaginary unit, and $d_{1m}(t)$ is the RDOA between the $m$-th microphone and the reference ($m=1$) microphone at time $t$. 
As a result, neglecting any phase distortion due to noise,
\begin{align}
    \mathbf{\Lambda}(t,f) 
    &\approx -c\left(2 \pi f\right)^{-1}\arg
    \left[H_1^{\ast}(t,f)\mathbf{H}_{2:M}(t,f)\right]\\
    &\approx \left[d_{12}(t), \dots, d_{1M}(t)\right]^\T, \label{eq:rdoa}
\end{align}
that is, $\mathbf{\Lambda}$ also approximates the RDOAs in a similar manner to $\mathbf{V}$.

In practice, NIPD is, admittedly, considerably noisier than EPV, as the latter benefited from the noise suppressing properties of the eigendecomposition. However, the use of $\mathbf{X}$ for $\mathbf{\Lambda}$ in contrast to $\mathbf{U}$ for $\mathbf{V}$ in the original SALSA significantly reduces the computational time, as the singular value decomposition required to compute $\mathbf{U}$ from $\mathbf{X}$ has a complexity of $\mathcal{O}(M^3F)$ per time frame \cite{golub2013matcal}. 

Compared to the \textsc{MelSpecGCC}, SALSA-Lite is also faster to compute. More importantly, SALSA-Lite has the exact TF alignment between the multichannel spectrograms and the NIPD in linear-frequency scale, which is crucial for resolving overlapping sound events. In addition, this TF alignment is arguably more suitable for CNN-based models, of which the kernels learn patterns from small multichannel TF patches of the image-like SELD input features. 

As with SALSA, the values of $\mathbf{\Lambda}$ for all TF bins above the aliasing frequency are zeroed. For efficiency, the magnitude and coherence tests are not used for SALSA-Lite\footnote{
Magnitude test is more useful when more frequency bands are included in the spatial features, i.e. high cut-off frequency. Coherence test requires heavy computation due to the need of computing eigenvalues~\cite{tho2021salsa}.}. We also experimented with an ablation variant of SALSA-Lite called SALSA-IPD to evaluate the usefulness of the frequency normalization. SALSA-IPD uses frequency-dependent IPD, which is only normalized by $(-2\pi)$, instead of $(-2\pi f / c)$ as in NIPD.

\section{Experimental settings}
\label{sec:exp}

We compared the performances of SELD models trained on the proposed SALSA-Lite and SALSA-IPD features with those trained on SALSA and \textsc{MelSpecGCC} features. 
We used the same experimental settings, such as network architecture, data augmentation and hyperparameters, for all features\footnote{See \href{https://github.com/thomeou/SALSA-Lite}{http://github.com/thomeou/SALSA-Lite} for details.}.

\subsection{Network architecture for SELD}
\label{sec:net}

\begin{figure}[t]
\centering
\includegraphics[width=0.85\columnwidth]{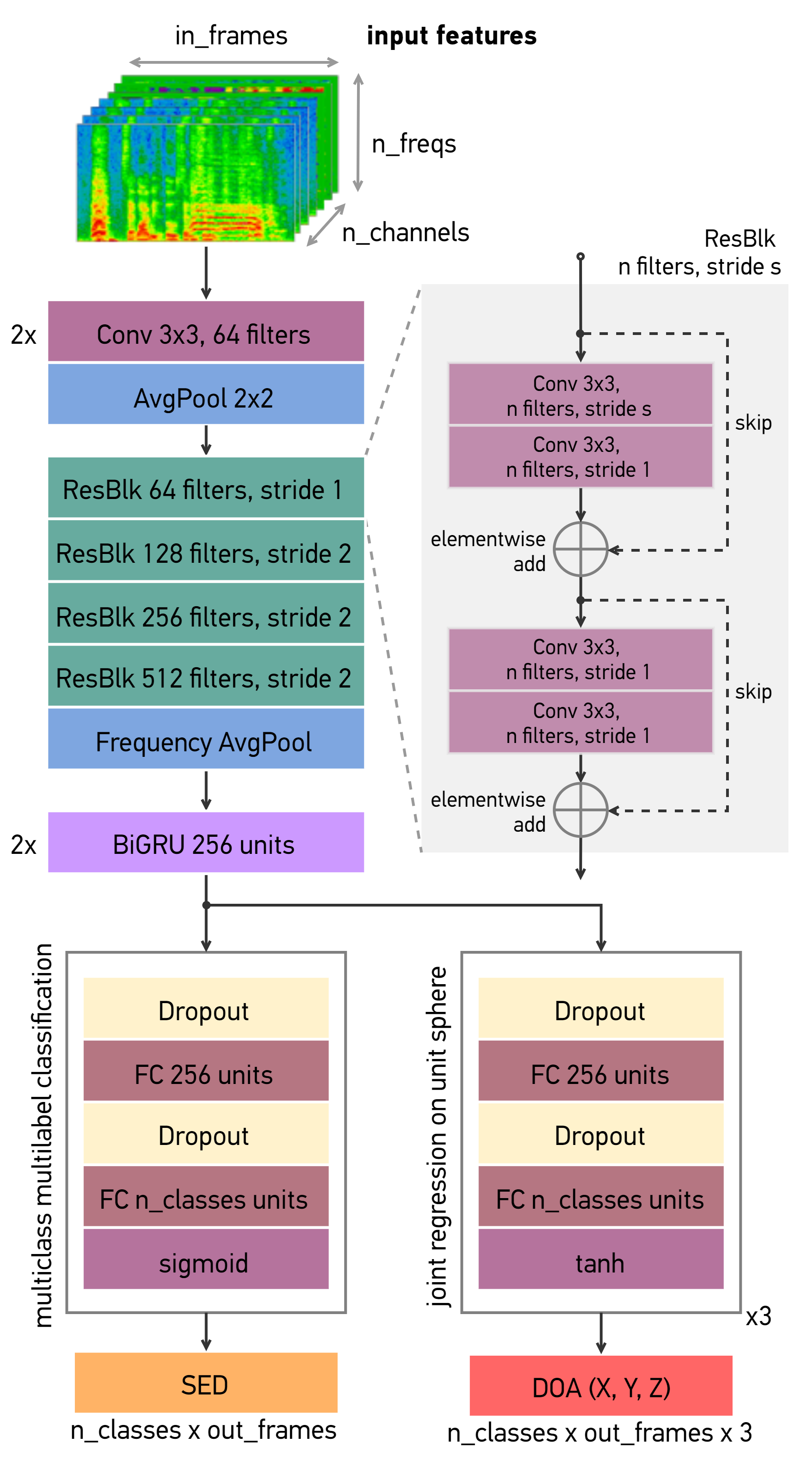}
\caption{Block diagram of the SELD network \cite{tho2021salsa}.}
\label{fig:seldnet}
\end{figure}

Fig.~\ref{fig:seldnet} shows the SELD network architecture used for all the experiments in this paper. The CRNN network consists of a convolutional backbone that is based on ResNet22 for audio tagging~\cite{kong2019panns}, a two-layer BiGRU, and fully connected (FC) layers. The number of input channels in the first convolutional layer is set to the number of channels of each feature under study. The SED branch is formulated as a multilabel multiclass classification and the DOAE branch as a three-dimensional Cartesian regression. During inference, sound classes whose probabilities are above the SED threshold are considered active classes. The DOAs corresponding to these classes are selected accordingly.

\subsection{Data augmentation}
\label{sec:dataaug}

To mitigate the problem of small datasets in SELD, we applied three data augmentation techniques to all the features: channel swapping (CS)~\cite{wang2021fouraug}, random cutout (RC)~\cite{zhong2017random, park2019specaugment}, and frequency shifting (FS)~\cite{tho2021salsa}. 
These augmentation techniques can be applied to the data in STFT domain during training. 
Only CS changes the ground truth, while RC and FS do not alter the target labels. In CS, there are \num{8} ways to swap channels for the MIC format~\cite{wang2021fouraug}. The spectrograms, GCC-PHAT, EPV, IPD, NIPD and target labels are altered accordingly when the channels are swapped. In RC, either random cutout~\cite{zhong2017random} or TF masking via SpecAugment~\cite{park2019specaugment} was applied on all channels of the input features. Random cutout produces a rectangular mask on the spectrograms while SpecAugment produces a cross-shaped mask. 
For FS, we randomly shifted all the input features up or down along the frequency dimension by up to \num{10} bands~\cite{tho2021salsa}. For \textsc{MelSpecGCC} features, the GCC-PHAT channels are not shifted.

\subsection{Dataset}
\label{sec:dataset}

The development split of the TNSSE 2021 dataset~\cite{politis2021dcasedataset} was used for our experiments. 
The MIC-format dataset consists of \num{400}, \num{100}, and \num{100} one-minute four-channel audio clips, sampled at \SI{24}{\kilo\hertz}, for training, validation, and testing, respectively. There are \num{12} sound classes. The azimuth and elevation ranges of both datasets are $[\SI{-180}{\degree}, \SI{180}{\degree})$ and $[\SI{-45}{\degree}, \SI{45}{\degree}]$, respectively. The validation set was used for model selection while the test set was used for evaluation. Since the distance between microphones of the provided microphone array corresponds to an aliasing frequency of \SI{2}{\kilo\hertz}, we computed the IPD and NIPD between \SI{50}{\hertz} and \SI{2}{\kilo\hertz}. For SALSA feature, the EPV was computed between \SI{50}{\hertz} and \SI{4}{\kilo\hertz} as per~\cite{tho2021salsa}\footnote{We obtained lower SELD performance when IPD and NIPD were computed between \SI{50}{\hertz} and \SI{4}{\kilo\hertz}}. Even though the array in \cite{politis2021dcasedataset} was mounted on an acoustically-hard spherical baffle, we found that the farfield array model in Section~\ref{subsec:salsa-lite} is sufficient to extract the spatial cues for the spherical array~\cite{tho2021salsa}. 

\subsection{Evaluation metrics}
\label{sec:metrics}

The official 2021 DCASE Challenge metrics~\cite{politis2020overview} were used for evaluation. 
These SELD evaluation metrics consist of four metrics: location-dependent error rate ($\text{ER}_{\le \SI{20}{\degree}}$) and F1 score ($\text{F}_{\le \SI{20}{\degree}}$) for SED; and class-dependent localization error ($\text{LE}_\text{CD}$), and localization recall ($\text{LR}_\text{CD}$) for DOAE. We also reported an SELD error which was computed as 
\begin{equation}
    \mathcal{E}_\textsc{seld} = \dfrac{1}{4}\left[\text{ER}_{\le \SI{20}{\degree}} + (1-\text{F}_{\le \SI{20}{\degree}}) + \dfrac{\text{LE}_\textsc{cd}}{\SI{180}{\degree}} + (1-\text{LR}_\textsc{cd})\right],
\end{equation}
to aggregate all four metrics. A good SELD system should have low ER, high F1, low DE, high FR, and low $\mathcal{E}_\textsc{seld}$. 

\subsection{Hyperparameters and training procedure}

We used an STFT window length of \SI{512}{ samples}, hop length of \SI{300}{samples}, Hann window, \num{512} FFT points, and \num{128} mel bands. 
A cutoff frequency of \SI{9}{\kilo\hertz} was used to compute all the feature. As a result, the frequency dimension has 192 bins for SALSA, SALSA-lite and SALSA-IPD features. \num{8}-second audio chunks were used for training while the full \num{60}-second audio clips were used for inference. 
The Adam optimizer \cite{Kingma2014Adam:Optimization} was used, with the learning rate initially set to \num{3e-4} and linearly decreased to \num{e-4} over last \num{15} epochs of the total \num{50} training epochs. A threshold of \num{0.3} was selected through hyper-parameter search using the validation split to binarize active class predictions in the SED outputs. 

\section{Experimental results and discussion}
\label{sec:results}

\begin{table}[t]
    \setlength\tabcolsep{3pt}
    \centering
    \caption{SELD performances for different features.}
    \noindent\begin{tabularx}{\columnwidth}{Xcrrrrr}
    \toprule 
        Feature & Aug.
        & $\text{ER}_{\le \SI{20}{\degree}}$ 
        & $\text{F}_{\le \SI{20}{\degree}}$
        & $\text{LE}_\text{CD}$
        & $\text{LR}_\text{CD}$ 
        & $\mathcal{E}_\text{SELD}$ \\
    \midrule
    \textsc{MelSpecGCC} & N          & 0.660 & 0.455 & 21.1\si{\degree} & 0.521 & 0.450 \\ 
        SALSA    & N           & 0.528 & 0.601 & \bf{15.9\si{\degree}} & 0.644 & 0.343 \\
        SALSA-IPD  & N         & 0.542 & 0.576 & 17.5\si{\degree} & 0.635 & 0.357 \\
        SALSA-Lite & N         & \bf{0.512} & \bf{0.609} & 16.9\si{\degree} & \bf{0.657} & \bf{0.335} \\
    \midrule
        \textsc{MelSpecGCC}    & Y      & 0.507 & 0.614 & 17.9\si{\degree} & 0.679 & 0.328 \\
        SALSA    & Y           & \bf{0.408} & \bf{0.715} & 12.6\si{\degree} & \bf{0.728} & \bf{0.259} \\
        SALSA-IPD  & Y         & 0.415 & 0.703 & 12.4\si{\degree} & 0.701 & 0.270 \\
        SALSA-Lite   & Y       & 0.409 & 0.707 & \bf{12.3\si{\degree}} & 0.716 & 0.264 \\
    \bottomrule
    \end{tabularx}
    \label{tab:seld_results}
\end{table}

\Cref{tab:seld_results} shows performances of all examined features with and without data augmentation. The results across all the metrics clearly show that data augmentation greatly benefits all of the features, of which \textsc{MelSpecGCC} received the largest gain. 
On average, data augmentation improves the \ER, \Fone, \LE, and \LR metrics by \SI{22}{\percent},  \SI{25}{\percent}, \SI{4}{\degree}, and \SI{15}{\percent}, respectively. 
With and without data augmentation, SALSA-based features outperformed \textsc{MelSpecGCC} feature by a large margin across all the evaluation metrics. This result clearly shows the advantages of SALSA-based features of having exact TF alignment between the spectral and spatial features over the simple stacking of spectrograms and GCC-PHAT spectra as per \textsc{MelSpecGCC}. In addition, this exact TF mapping might be more suitable for the learning of CNNs, as the filters can more conveniently learn the SELD patterns from sections of the image-like input features. Furthermore, these results show that EPV, IPD and NIPD provide better spatial cues for SELD than \gccphat spectra. 

Without data augmentation, SALSA-Lite achieved the highest overall performance, followed by SALSA and SALSA-IPD, respectively. With data augmentation, although SALSA was the best-performing feature, SALSA-Lite and SALSA-IPD only performed slightly worse. 
For a \num{60}-second audio clip with \num{4} input channels, using a machine with a 10-core Intel i9-7900X CPU, SALSA-Lite and SALSA-IPD take only \SI{0.30}{\second} on average for feature computation, 9 and 30 times faster than \textsc{MelSpecGCC} (\SI{2.90}{\second}) and SALSA (\SI{9.45}{\second}) respectively. The small performance gap between SALSA and SALSA-Lite together with the huge speedup factor of SALSA-Lite show that SALSA-Lite is a more attractive feature for SELD applications that requires both fast computation and high performance, such as real-time SELD. 

With and without data augmentation, SALSA-Lite outperformed SALSA-IPD, which shows that the simple frequency-normalization trick for NIPD is effective. As shown in \eqref{eq:rdoa}, NIPD channels in SALSA-Lite are theoretically frequency-invariant for single-source TF bins thus any difference in values of the NIPD channels across the frequency axis can be attributed to noise. On the other hand, unnormalized IPD channels in SALSA-IPD has frequency-dependencies in addition to noise. The increased learning burden due to the lack of normalization likely contributed to the performance gap. Regardless, the small performance gap between SALSA-Lite and SALSA-IPD indicated that CNNs are able to learn useful spatial cues from the frequency-dependent IPD even without normalization. 

\begin{table}[t]
    \setlength\tabcolsep{3pt}
    \centering
    \caption{Effect of spatial aliasing on SALSA-Lite and SALSA-IPD.}
    \noindent\begin{tabularx}{\columnwidth}{Xcrrrrr}
    \toprule 
        Feature 
        & Cut-off
        & $\text{ER}_{\le \SI{20}{\degree}}$ 
        & $\text{F}_{\le \SI{20}{\degree}}$
        & $\text{LE}_\text{CD}$
        & $\text{LR}_\text{CD}$ 
        & $\mathcal{E}_\text{SELD}$ \\
    \midrule
    SALSA-IPD &
        \SI{2}{\kilo\hertz}            & \bf{0.415} & \bf{0.703} & \bf{12.4\si{\degree}} & \bf{0.701} & \bf{0.270} \\
        & \SI{9}{\kilo\hertz}            & 0.434 & 0.690 & \bf{12.4\si{\degree}} & 0.699 & 0.279 \\
    \midrule
    SALSA-Lite &
        \SI{2}{\kilo\hertz}            & \bf{0.409} & \bf{0.707} & \bf{12.3\si{\degree}} & \bf{0.716} & \bf{0.264} \\
        & \SI{9}{\kilo\hertz}            & 0.423 & 0.699 & 12.6\si{\degree} & 0.714 & 0.270 \\
    \bottomrule
    \end{tabularx}
    \label{tab:aliasing}
\end{table}

We report the performance of SALSA-Lite and SALSA-IPD with upper cutoff frequency for NIPD and IPD at \SI{2}{\kilo\hertz} and \SI{9}{\kilo\hertz} (full-band) in \Cref{tab:aliasing} to examine the effect of spatial aliasing on SELD performance. 
For both features, \SI{2}{\kilo\hertz} cutoff performed slightly better than full band. However, similar to SALSA feature~\cite{tho2021salsa}, SALSA-Lite and SALSA-IPD are only mildly affected by spatial aliasing. These results agree with the finding in \cite{dmochowski2009spatialaliasing}, where broadband signals were shown to not affected by spatial aliasing unless they contain strong harmonic components. 

\begin{table}[t] \small
    {\centering
    \caption{SELD performances of state-of-the-art systems and SALSA-Lite models on the test split of TNSSE 2021 dataset.}
    \setlength{\tabcolsep}{3pt}
    \footnotesize
    \noindent\begin{tabularx}{\columnwidth}{Xlrrrrr}
    \toprule 
        System (\# params)
        & Format
        & $\text{ER}_{\le \SI{20}{\degree}}$ 
        & $\text{F}_{\le \SI{20}{\degree}}$
        & $\text{LE}_\text{CD}$
        & $\text{LR}_\text{CD}$ \\
    \midrule
        DCASE'21 baseline (0.5M) ~\cite{politis2021dcasedataset}
            & FOA & 0.73\hphantom{0} & 0.307 & 24.5\si{\degree} & 0.448 \\
            & MIC & 0.74\hphantom{0} & 0.247 & 30.9\si{\degree} & 0.382 \\
        \midrule
        (\#1) Shimada et al.~(42M)~\cite{shimada2021dcasetop}$\star$
            & FOA & 0.43\hphantom{0} & 0.699 & \bf{11.1\si{\degree}} & 0.732 \\
        (\#2) Nguyen et at.~(107M)~\cite{tho2021salsachallenge}$\star$
        & FOA & \bf{0.37\hphantom{0}} & \bf{0.737} & 11.2\si{\degree} & \bf{0.741} \\
        (\#4) Lee et at.~(27M)~\cite{lee2021crossmodalseld}$\star$
            & FOA & 0.46\hphantom{0} & 0.609 & 14.4\si{\degree} & 0.733 \\
    \midrule
        SALSA-Lite (14M)
            & MIC & 0.409 & 0.707 & 12.3\si{\degree} & 0.716\\
    \bottomrule
    \end{tabularx}
    \label{tab:sota_2021_dev}
    } 
    $\star$ denotes an ensemble model.
    The bracket denotes DCASE 2021 Task 3 ranking on the evaluation split.
\end{table}

\Cref{tab:sota_2021_dev} shows the performances on the test split of the TNSSE 2021 dataset of state-of-the-art systems, all of which are ensemble models, and our single-CRNN model trained on SALSA-Lite. Since there is a severe lack of SELD systems developed for MIC format, we also included SELD systems developed for FOA format. The model trained on SALSA-Lite feature significantly outperformed the DCASE baseline for MIC format. Even though our model is only a simple CRNN, it performed better than the highest-ranked ensemble from the 2021 DCASE Challenge \cite{shimada2021dcasetop} in terms of \ER and \Fone, and only slightly worse in terms of \LE and \LR. 
The results show that the proposed SALSA-Lite features for MIC formats are effective for SELD.

\section{Conclusions}
\label{sec:conclusion}

In conclusion, the proposed SALSA-Lite feature addresses the lack of fast and effective feature for SELD using microphone array. SALSA stands for Spatial cue-Augmented Log-SpectrogAm. It is simple to compute, and significantly faster than \textsc{MelSpecGCC} and full SALSA features. It achieves better performance than \textsc{MelSpecGCC} and on-par performance with SALSA. It is less affected by spatial aliasing. Simple CRNN models trained on SALSA-Lite feature achieve comparative performance compared to many state-of-the-art SELD systems. In addition, there are many effective data augmentation techniques that can be applied on SALSA-Lite on the fly during training to further improve model performance.  




\renewcommand{\bibsection}{\section{References}}
\setlength{\bibsep}{9pt plus 2pt minus 9pt}
\bibliographystyle{IEEEbib}
\bibliography{references}

\begin{thebibliography}{10}
\providecommand{\url}[1]{#1}
\def\UrlFont{\rmfamily}
\providecommand{\newblock}{\relax}
\providecommand{\bibinfo}[2]{#2}
\providecommand\BIBentrySTDinterwordspacing{\spaceskip=0pt\relax}
\providecommand\BIBentryALTinterwordstretchfactor{4}
\providecommand\BIBentryALTinterwordspacing{\spaceskip=\fontdimen2\font plus
\BIBentryALTinterwordstretchfactor\fontdimen3\font minus
  \fontdimen4\font\relax}
\providecommand\BIBforeignlanguage[2]{{%
\expandafter\ifx\csname l@#1\endcsname\relax
\typeout{** WARNING: IEEEtran.bst: No hyphenation pattern has been}%
\typeout{** loaded for the language `#1'. Using the pattern for}%
\typeout{** the default language instead.}%
\else
\language=\csname l@#1\endcsname
\fi
#2}}

\bibitem{adavanne2019seld}
S.~Adavanne, A.~Politis, J.~Nikunen, and T.~Virtanen, ``{Sound Event
  Localization and Detection of Overlapping Sources Using Convolutional
  Recurrent Neural Networks},'' \emph{\protect\JournalTitle{IEEE Journal of
  Selected Topics in Signal Processing}}, vol.~13, no.~1, pp. 34--48, 2019.

\bibitem{cao2019twostage}
Y.~Cao, Q.~Kong, T.~Iqbal, F.~An, W.~Wang, and M.~D. Plumbley, ``{Polyphonic
  Sound Event Detection and Localization using a Two-Stage Strategy},'' in
  \emph{\protect\JournalTitle{Proceedings of the Detection and Classification
  of Acoustic Scenes and Events 2019 Workshop (DCASE2019)}}, 2019, pp. 30--34.

\bibitem{tho2020smn}
T.~N.~T. Nguyen, D.~L. Jones, and W.-S. Gan, ``{A Sequence Matching Network for
  Polyphonic Sound Event Localization and Detection},'' in
  \emph{\protect\JournalTitle{International Conference on Acoustics, Speech and
  Signal Processing (ICASSP)}}, 2020, pp. 71--75.

\bibitem{tho2021gennet}
T.~N.~T. Nguyen, N.~K. Nguyen, H.~Phan, L.~Pham, K.~Ooi, D.~L. Jones, and W.-S.
  Gan, ``{A General Network Architecture for Sound Event Localization and
  Detection Using Transfer Learning and Recurrent Neural Network},'' in
  \emph{\protect\JournalTitle{ICASSP}}, 2021, pp. 1--5.

\bibitem{cao2021ienv2}
Y.~Cao, T.~Iqbal, Q.~Kong, F.~An, W.~Wang, and M.~D. Plumbley, ``{An Improved
  Event-Independent Network for Polyphonic Sound Event Localization and
  Detection},'' in \emph{\protect\JournalTitle{International Conference on
  Acoustics, Speech and Signal Processing (ICASSP)}}, 2021, pp. 885--889.

\bibitem{shimada2021accdoa}
K.~Shimada, Y.~Koyama, N.~Takahashi, S.~Takahashi, and Y.~Mitsufuji, ``{ACCDOA:
  Activity-Coupled Cartesian Direction of Arrival Representation for Sound
  Event Localization And Detection},'' in
  \emph{\protect\JournalTitle{International Conference on Acoustics, Speech and
  Signal Processing (ICASSP)}}, 2021, pp. 915--919.

\bibitem{takahashi2021d3net}
\BIBentryALTinterwordspacing
N.~Takahashi and Y.~Mitsufuji, ``{Densely connected multidilated convolutional
  networks for dense prediction tasks},'' in
  \emph{\protect\JournalTitle{Proceedings of the IEEE/CVF Conference on
  Computer Vision and Pattern Recognition (CVPR)}}, 2021, pp. 993--1002.

\bibitem{politis2020dcasedataset}
A.~Politis, S.~Adavanne, and T.~Virtanen, ``{A Dataset of Reverberant Spatial
  Sound Scenes with Moving Sources for Sound Event Localization and
  Detection},'' in \emph{\protect\JournalTitle{Proceedings of the Detection and
  Classification of Acoustic Scenes and Events 2020 Workshop (DCASE2020)}},
  2020, pp. 165--169.

\bibitem{politis2021dcasedataset}
\BIBentryALTinterwordspacing
A.~Politis, S.~Adavanne, D.~Krause, A.~Deleforge, P.~Srivastava, and
  T.~Virtanen, ``{A Dataset of Dynamic Reverberant Sound Scenes with
  Directional Interferers for Sound Event Localization and Detection},''
  \emph{\protect\JournalTitle{arXiv:2106.06999}}, 2021.
\BIBentrySTDinterwordspacing

\bibitem{wang2021fouraug}
\BIBentryALTinterwordspacing
Q.~Wang, J.~Du, H.-X. Wu, J.~Pan, F.~Ma, and C.-H. Lee, ``{A Four-Stage Data
  Augmentation Approach to ResNet-Conformer Based Acoustic Modeling for Sound
  Event Localization and Detection},''
  \emph{\protect\JournalTitle{arXiv:2101.02919}}, 2021.
\BIBentrySTDinterwordspacing

\bibitem{he2021sounddet}
\BIBentryALTinterwordspacing
Y.~He, N.~Trigoni, and A.~Markham, ``{SoundDet: Polyphonic Moving Sound Event
  Detection and Localization from Raw Waveform},'' in
  \emph{\protect\JournalTitle{International Conference on Machine Learning}},
  2021.

\bibitem{tho2021salsa}
T.~N.~T. Nguyen, K.~N. Watcharasupat, N.~K. Nguyen, D.~L. Jones, and W.-S. Gan,
  ``{SALSA: Spatial cue-augmented log spectrogram for polyphonic sound event
  localization and detection},''
  \emph{\protect\JournalTitle{arXiv:2110.00275}}, 2021.

\bibitem{araki2007bss}
\BIBentryALTinterwordspacing
S.~Araki, H.~Sawada, R.~Mukai, and S.~Makino, ``{Underdetermined Blind Sparse
  Source Separation for Arbitrarily Arranged Multiple Sensors},''
  \emph{\protect\JournalTitle{Signal Processing}}, vol.~87, no.~8, p.
  1833–1847, 2007.
\BIBentrySTDinterwordspacing

\bibitem{traa2013bssipd}
J.~Traa and P.~Smaragdis, ``{Blind multi-channel source separation by
  circular-linear statistical modeling of phase differences},'' in
  \emph{\protect\JournalTitle{2013 IEEE International Conference on Acoustics,
  Speech and Signal Processing}}, 2013, pp. 4320--4324.

\bibitem{wang2018multichannelclustering}
Z.-Q. Wang, J.~Le~Roux, and J.~R. Hershey, ``{Multi-Channel Deep Clustering:
  Discriminative Spectral and Spatial Embeddings for Speaker-Independent Speech
  Separation},'' in \emph{\protect\JournalTitle{2018 IEEE International
  Conference on Acoustics, Speech and Signal Processing (ICASSP)}}, 2018, pp.
  1--5.

\bibitem{zhang2010twomicdoa}
W.~Zhang and B.~D. Rao, ``{A Two Microphone-Based Approach for Source
  Localization of Multiple Speech Sources},'' \emph{\protect\JournalTitle{IEEE
  Transactions on Audio, Speech, and Language Processing}}, vol.~18, no.~8, pp.
  1913--1928, 2010.

\bibitem{pak2019ipdenhancementdnn}
J.~Pak and J.~W. Shin, ``{Sound Localization Based on Phase Difference
  Enhancement Using Deep Neural Networks},''
  \emph{\protect\JournalTitle{IEEE/ACM Transactions on Audio, Speech, and
  Language Processing}}, vol.~27, no.~8, pp. 1335--1345, 2019.

\bibitem{schmidt1986multiple}
R.~Schmidt, ``{Multiple emitter location and signal parameter estimation},''
  \emph{\protect\JournalTitle{IEEE Trans. Antennas Propag.}}, vol.~34, no.~3,
  pp. 276--280, 1986.

\bibitem{capon1969mvdr}
J.~Capon, ``{High-Resolution Frequency-Wavenumber Spectrum Analysis},'' in
  \emph{\protect\JournalTitle{Proceedings of the IEEE}}, 1969, pp. 1408--1418.

\bibitem{golub2013matcal}
G.~H. Golub and C.~F. Van~Loan, \emph{\protect\JournalTitle{{Matrix
  Computations}}}.\hskip 1em plus 0.5em minus 0.4em\relax Baltimore: Johns
  Hopkins University Press, 2013.

\bibitem{kong2019panns}
\BIBentryALTinterwordspacing
Q.~Kong, Y.~Cao, T.~Iqbal, Y.~Wang, W.~Wang, and M.~D. Plumbley, ``{PANNs:
  Large-Scale Pretrained Audio Neural Networks for Audio Pattern
  Recognition},'' \emph{\protect\JournalTitle{IEEE/ACM Transactions on Audio,
  Speech, and Language Processing}}, vol.~28, pp. 2880--2894, 2020.
\BIBentrySTDinterwordspacing

\bibitem{zhong2017random}
Z.~Zhong, L.~Zheng, G.~Kang, S.~Li, and Y.~Yang, ``{Random erasing data
  augmentation},'' in \emph{\protect\JournalTitle{Proceedings of the 34th AAAI
  Conference on Artificial Intelligence}}, 2020, pp. 13\,001--13\,008.

\bibitem{park2019specaugment}
D.~S. Park, W.~Chan, Y.~Zhang, C.~C. Chiu, B.~Zoph, E.~D. Cubuk, and Q.~V. Le,
  ``{SpecAugment: A simple data augmentation method for automatic speech
  recognition},'' in \emph{\protect\JournalTitle{Proceedings of the Annual
  Conference of the International Speech Communication Association}}, 2019, pp.
  2613--2617.

\bibitem{politis2020overview}
A.~Politis, A.~Mesaros, S.~Adavanne, T.~Heittola, and T.~Virtanen, ``{Overview
  and Evaluation of Sound Event Localization and Detection in DCASE 2019},''
  \emph{\protect\JournalTitle{IEEE/ACM Transactions on Audio, Speech, and
  Language Processing}}, vol.~29, pp. 684--698, 2020.

\bibitem{Kingma2014Adam:Optimization}
\BIBentryALTinterwordspacing
D.~P. Kingma and J.~Ba, ``{Adam: A Method for Stochastic Optimization},'' in
  \emph{\protect\JournalTitle{Conference Track Proceedings of the 3rd
  International Conference on Learning Representations}}, 2014.

\bibitem{dmochowski2009spatialaliasing}
J.~Dmochowski, J.~Benesty, and S.~Affes, ``{On Spatial Aliasing in Microphone
  Arrays},'' \emph{\protect\JournalTitle{IEEE Transactions on Signal
  Processing}}, vol.~57, no.~4, pp. 1383--1395, 2009.

\bibitem{shimada2021dcasetop}
K.~Shimada, N.~Takahashi, Y.~Koyama, S.~Takahashi, E.~Tsunoo, M.~Takahashi, and
  Y.~Mitsufuji, ``{Ensemble of ACCDOA- and EINV2-based Systems with D3Nets and
  Impulse Response Simulation for Sound Event Localization and Detection},''
  DCASE2021 Challenge, Tech. Rep., 2021.

\bibitem{tho2021salsachallenge}
T.~N.~T. Nguyen, K.~Watcharasupat, N.~K. Nguyen, D.~L. Jones, and W.-S. Gan,
  ``{DCASE 2021 Task 3: Spectrotemporally-aligned Features for Polyphonic Sound
  Event Localization and Detection},'' DCASE2021 Challenge, Tech. Rep., 2021.

\bibitem{lee2021crossmodalseld}
S.-h. Lee, J.-w. Hwang, S.-b. Seo, and H.-m. Park, ``{Sound Event Localization
  and Detection Using Cross-modal Attention and Parameter Sharing for DCASE2021
  Challenge},'' DCASE2021 Challenge, Tech. Rep., 2021.

\end{thebibliography}

\end{document}